\documentclass[twocolumn,showpacs,prl,superscriptaddress]{revtex4}          
\def\sNN{\mbox{$\sqrt{s_{_{NN}}}$}}   
\newcommand{ \be }{\begin{equation}}       
\newcommand{ \ee }{\end{equation}}       
\newcommand{ \bea }{\begin{eqnarray}}       
\newcommand{ \eea }{\end{eqnarray}}

\newcommand{ \mean }[1]{\left\langle #1 \right\rangle}   
\newcommand{ \etal }{{\it et al.}}   
\usepackage{color}

\usepackage{graphicx} 
\usepackage{fixmath} 
\usepackage{dcolumn} 
\usepackage{bm} 
\usepackage{multirow}
 
\begin{document}          
\title{       
\begin{flushright}  
{\small \sl version 14.2,  
\today \\  
 } 
\end{flushright} 
System-size independence of directed flow at the Relativistic Heavy-Ion Collider
} 
 
 
\affiliation{Argonne National Laboratory, Argonne, Illinois 60439, USA}
\affiliation{University of Birmingham, Birmingham, United Kingdom}
\affiliation{Brookhaven National Laboratory, Upton, New York 11973, USA}
\affiliation{California Institute of Technology, Pasadena, California 91125, USA}
\affiliation{University of California, Berkeley, California 94720, USA}
\affiliation{University of California, Davis, California 95616, USA}
\affiliation{University of California, Los Angeles, California 90095, USA}
\affiliation{Universidade Estadual de Campinas, Sao Paulo, Brazil}
\affiliation{Carnegie Mellon University, Pittsburgh, Pennsylvania 15213, USA}
\affiliation{University of Illinois at Chicago, Chicago, Illinois 60607, USA}
\affiliation{Creighton University, Omaha, Nebraska 68178, USA}
\affiliation{Nuclear Physics Institute AS CR, 250 68 \v{R}e\v{z}/Prague, Czech Republic}
\affiliation{Laboratory for High Energy (JINR), Dubna, Russia}
\affiliation{Particle Physics Laboratory (JINR), Dubna, Russia}
\affiliation{University of Frankfurt, Frankfurt, Germany}
\affiliation{Institute of Physics, Bhubaneswar 751005, India}
\affiliation{Indian Institute of Technology, Mumbai, India}
\affiliation{Indiana University, Bloomington, Indiana 47408, USA}
\affiliation{Institut de Recherches Subatomiques, Strasbourg, France}
\affiliation{University of Jammu, Jammu 180001, India}
\affiliation{Kent State University, Kent, Ohio 44242, USA}
\affiliation{University of Kentucky, Lexington, Kentucky, 40506-0055, USA}
\affiliation{Institute of Modern Physics, Lanzhou, China}
\affiliation{Lawrence Berkeley National Laboratory, Berkeley, California 94720, USA}
\affiliation{Massachusetts Institute of Technology, Cambridge, MA 02139-4307, USA}
\affiliation{Max-Planck-Institut f\"ur Physik, Munich, Germany}
\affiliation{Michigan State University, East Lansing, Michigan 48824, USA}
\affiliation{Moscow Engineering Physics Institute, Moscow Russia}
\affiliation{City College of New York, New York City, New York 10031, USA}
\affiliation{NIKHEF and Utrecht University, Amsterdam, The Netherlands}
\affiliation{Ohio State University, Columbus, Ohio 43210, USA}
\affiliation{Panjab University, Chandigarh 160014, India}
\affiliation{Pennsylvania State University, University Park, Pennsylvania 16802, USA}
\affiliation{Institute of High Energy Physics, Protvino, Russia}
\affiliation{Purdue University, West Lafayette, Indiana 47907, USA}
\affiliation{Pusan National University, Pusan, Republic of Korea}
\affiliation{University of Rajasthan, Jaipur 302004, India}
\affiliation{Rice University, Houston, Texas 77251, USA}
\affiliation{Universidade de Sao Paulo, Sao Paulo, Brazil}
\affiliation{University of Science \& Technology of China, Hefei 230026, China}
\affiliation{Shanghai Institute of Applied Physics, Shanghai 201800, China}
\affiliation{SUBATECH, Nantes, France}
\affiliation{Texas A\&M University, College Station, Texas 77843, USA}
\affiliation{University of Texas, Austin, Texas 78712, USA}
\affiliation{Tsinghua University, Beijing 100084, China}
\affiliation{Valparaiso University, Valparaiso, Indiana 46383, USA}
\affiliation{Variable Energy Cyclotron Centre, Kolkata 700064, India}
\affiliation{Warsaw University of Technology, Warsaw, Poland}
\affiliation{University of Washington, Seattle, Washington 98195, USA}
\affiliation{Wayne State University, Detroit, Michigan 48201, USA}
\affiliation{Institute of Particle Physics, CCNU (HZNU), Wuhan 430079, China}
\affiliation{Yale University, New Haven, Connecticut 06520, USA}
\affiliation{University of Zagreb, Zagreb, HR-10002, Croatia}

\author{B.~I.~Abelev}\affiliation{University of Illinois at Chicago, Chicago, Illinois 60607, USA}
\author{M.~M.~Aggarwal}\affiliation{Panjab University, Chandigarh 160014, India}
\author{Z.~Ahammed}\affiliation{Variable Energy Cyclotron Centre, Kolkata 700064, India}
\author{B.~D.~Anderson}\affiliation{Kent State University, Kent, Ohio 44242, USA}
\author{D.~Arkhipkin}\affiliation{Particle Physics Laboratory (JINR), Dubna, Russia}
\author{G.~S.~Averichev}\affiliation{Laboratory for High Energy (JINR), Dubna, Russia}
\author{Y.~Bai}\affiliation{NIKHEF and Utrecht University, Amsterdam, The Netherlands}
\author{J.~Balewski}\affiliation{Massachusetts Institute of Technology, Cambridge, MA 02139-4307, USA}
\author{O.~Barannikova}\affiliation{University of Illinois at Chicago, Chicago, Illinois 60607, USA}
\author{L.~S.~Barnby}\affiliation{University of Birmingham, Birmingham, United Kingdom}
% ...

\collaboration{STAR Collaboration}\noaffiliation

\begin{abstract}     
We measure directed flow ($v_1$) for charged particles in Au+Au and Cu+Cu 
collisions at $\sNN =$ 200 GeV and 62.4 GeV, as a function of pseudorapidity ($\eta$), transverse 
momentum ($p_t$) and collision centrality, based on data from the STAR experiment.  
We find that the directed flow depends on the incident energy but, 
contrary to all available model implementations, 
not on the size of the colliding system at a given centrality. 
We extend the validity of the limiting fragmentation concept
to $v_1$ in different collision systems, and 
investigate possible explanations for the observed sign change in $v_1(p_t)$.  
\end{abstract} 
 
  
\maketitle  

The heavy ion program at the Relativistic Heavy Ion Collider (RHIC) seeks to
understand the nature and dynamics of strongly-interacting matter
under extreme conditions.
It is widely expected that in collisions at RHIC, 
a new partonic phase of matter is created, 
sQGP, strongly interacting Quark Gluon Plasma~\cite{whitepapers}.
In particular, its bulk nature is revealed in strong {\em elliptic} flow, which in central
collisions approaches the predictions of ideal hydrodynamics, assuming system thermalization
on an extremely short timescale ($\sim 0.5$~fm/c)~\cite{Heinz}.
   However, the mechanism behind such rapid thermalization remains far from clear
and is under active theoretical study~\cite{Kovchegov,Mrowczynski,Muller}.
   This may be related to another novel phenomenon that could be relevant at RHIC --- 
saturation of the gluon distribution --- which characterizes the nuclear parton distribution 
prior to collision~\cite{Kharzeev}.
   Various theoretical approaches to connect collision
geometry, saturated gluon distributions, and the onset of bulk collective behavior
are being explored~\cite{Heinz}; more experimental input would guide these efforts.

Directed flow refers to collective sidewards deflection of particles and is characterized by  a {\it first}-order harmonic ($v_1$) of the Fourier expansion of particle's azimuthal distribution w.r.t. the reaction plane~\cite{Methods}.
At large $\eta$ (in the fragmentation region) the directed flow is believed to be 
generated during the nuclear passage time ($2R/\gamma \sim 0.1$~fm/c)~\cite{sorge,Herrmann}.
   It therefore probes the onset of bulk collective dynamics during thermalization,
providing valuable experimental guidance to models of the pre-equilibrium stage.
   In this Letter, we present multiple-differential measurements of $v_1$ for Au+Au 
and Cu+Cu collisions at $\sqrt{s_{NN}}=$200 and 62.4~GeV as a function of  
$\eta$, $p_t$, and collision centrality.
Here, we report an intriguing new universal scaling of the 
phenomenon with collision centrality.
Existing implementations of Boltzmann/cascade and hydrodynamic 
models are unable to explain the measured trends.

\begin{figure}[t]
  \includegraphics[width=0.50\textwidth]{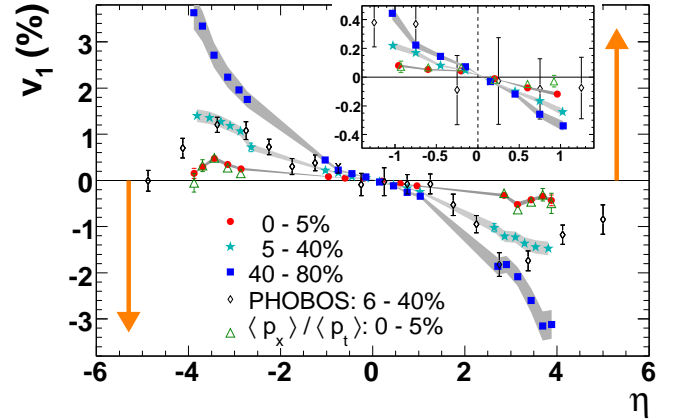}
  \caption{(color online) 
    Charged particle $v_1(\eta)$ for three centralities in Au+Au collisions at 200 GeV.
	      The arrows indicate the algebraic sign of $v_1$ 
            for spectator neutrons, and their positions on the $\eta$ axis correspond 
            to beam rapidity.
            The inset shows the mid-$\eta$ region in more detail. 
	   The error bars are statistical, and the shaded bands show systematic errors.
	      PHOBOS results~\cite{PHOBOSv1} are also shown for mid-central collisions.
   }
    \label{fig:AuAu200GeVEta}
\end{figure}

At RHIC energies, it is a challenge to measure $v_1$ accurately due to 
the relatively small signal and a potentially large systematic 
error arising from non-flow (azimuthal correlations not related to the 
reaction plane orientation).  In this work, the reaction plane was
determined from the sideward deflection of spectator neutrons
\cite{v1LowerEnergies,Herrmann} measured in the Shower Maximum 
Detectors (SMD) of the Zero Degree Calorimeters (ZDC)~\cite{v1ZDCSMD1, ZDC}.
The $v_1$ based on this quantity, denoted 
$v_1\{\text{ZDC-SMD}\}$~\cite{v1ZDCSMD1}, 
should have minimal contribution from non-flow effects due to the large 
$\eta$ gap between the spectator neutrons used to establish the reaction 
plane and the $\eta$ region where the measurements were performed.  

Charged particle tracks 
were reconstructed in STAR's main TPC~\cite{TPC-NIM} and forward TPCs~\cite{FTPC-NIM},
with pseudorapidity coverage $|\eta|<1.3$ and $2.5 < |\eta| < 4.0$, respectively.
The centrality definition 
(in which zero represents the most central collisions)
and track quality cuts are the same as in Ref.~\cite{Flow200GeV}.
This study is based on Au+Au samples of eight million events at 200 GeV, five million 
at 62.4 GeV, and Cu+Cu samples of twelve million events at 200 GeV, and eight million 
at 62.4 GeV. All were obtained with a minimum-bias trigger.  Systematic 
uncertainties on $v_1$ measurements are estimated to be within 
$10\%$ for the $\eta$ range studied.  
This limit is based on comparisons of $v_1\{\text{ZDC-SMD}\}$ and 
independent analysis methods~\cite{v1ZDCSMD1,Flow200GeV}, and we also make 
use of forward-backward symmetry to constrain estimates of systematic errors.  
Non-flow is not the dominant source of systematic uncertainty.  
More details about these errors can be found in Refs.~\cite{v1ZDCSMD1,Flow200GeV},

The resolution~\cite{Methods} of the first-order event plane reconstructed using 
the ZDC-SMDs is a crucial quantity for this analysis.
The magnitude of the event plane resolution, defined as  
$\mean{\cos(\Psi_{EP}-\Psi_{RP})}$~\cite{Methods}, 
increases with the spectator $v_1$ 
and the number of neutrons per event detected by the ZDC-SMDs.  The ZDC size is 
optimized for 200 GeV,  
and its acceptance for spectator neutrons decreases at lower energies 
due to spectator neutrons being emitted within a cone
whose apex angle increases with the inverse of the beam momentum.  
For the $30 - 60\%$ most central collisions, 
resolutions for 200 GeV Au+Au and Cu+Cu, and for
62.4 GeV Au+Au and Cu+Cu are about 0.4, 0.15, 0.15 and 0.04, respectively 
(more details are provided in Table 1 of Ref.~\cite{Gang_proceedings}.)
The $30 - 60\%$ centrality interval is the only region where
the ZDC-SMD event-plane resolution
can be reliably determined for all four systems.  

The charged-particle $v_1(\eta)$ is shown in Fig.~\ref{fig:AuAu200GeVEta} 
for Au+Au at $\sNN = 200$\,\,GeV in three centralities.
The inset shows, on expanded scales, the mid-$\eta$ region measured by the 
main TPC, where $v_1$ is resolvable below the 0.1\% level.    
Within the studied $\eta$ range, the sign of charged particle $v_1$ is opposite to
that of the spectators, and the $v_1$ magnitude increases from central to peripheral 
collisions.  
For 0-5\% centrality, 
the slope $dv_1/d\eta$ changes sign above the middle of the FTPC pseudorapidity 
acceptance, and our results 
agree with the pattern reported by PHOBOS over a broader $\eta$ 
range~\cite{PhobosQM,PHOBOSv1}.  

The ratio $\mean{p_x}/\mean{p_t}$ is shown in Fig.~\ref{fig:AuAu200GeVEta} for
the most central data (0 to 5\%), in comparison to $v_1$.  Here, $p_x$ refers 
to the in-plane component of a track's transverse momentum, a quantity 
commonly used prior to the 1990s~\cite{v1LowerEnergies}.
As elaborated below, there is interest in the behavior of both 
$v_1$ and $\langle p_x \rangle$ when $v_1(p_t)$ changes sign. 

\begin{figure}[t]
  \includegraphics[width=0.450\textwidth]{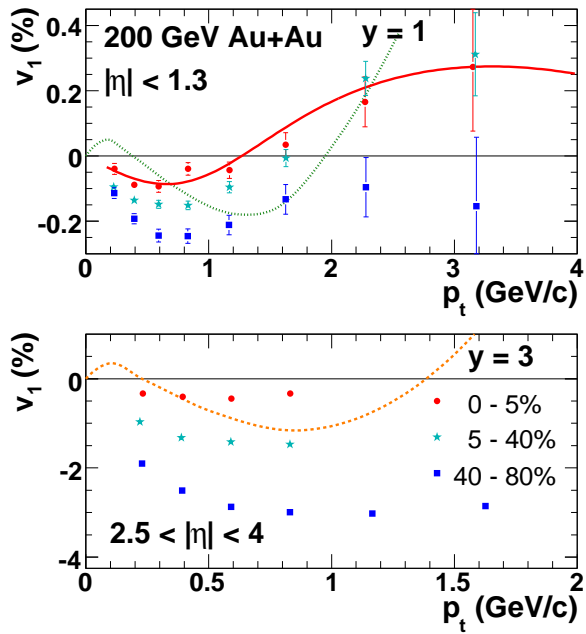}
  \caption{
    (color online) Charged particle $v_1(p_t)$ in 200 GeV Au+Au for three centralities. 
    The dashed and dotted curves are hydrodynamic 
    calculations for the labeled rapidities at impact parameter 
    6.8 fm ($15 - 25\%$ most central collisions). See the text for an explanation of the solid curve. 
    The plotted error bars are statistical, and systematic errors (see 
    Fig.~\ref{fig:AuAu200GeVEta}) are within 10\%.
}
    \label{fig:AuAu200GeVPt}
\end{figure}

To further examine $v_1$, the 200 GeV Au+Au data are divided into bins
of $p_t$ (Fig.~\ref{fig:AuAu200GeVPt}).
The upper and lower panels show results from the main TPC 
and the FTPCs, respectively.  In the main TPC, $v_1(p_t)$ crosses zero at 
$1 < p_t < 2$ GeV/$c$ for central and mid-central collisions.  
A zero-crossing behavior in $v_1(p_t)$ is necessarily exhibited by 
a hydrodynamic calculation in which $\mean{p_x}$, presumably imparted during 
the passing time of the initial-state nuclei, has been neglected and set equal to 
zero~\cite{Hydro_v1Pt}.  
Due to the poor momentum resolution of the FTPCs at higher $p_t$,
we cannot test the zero crossing at forward $\eta$.
It is noteworthy 
that the observed $\mean{p_x}$, presented in Fig.~\ref{fig:AuAu200GeVEta}, is 
far from negligible, which contradicts the assumptions used in the
hydrodynamic calculations. 

The observed $v_1(p_t)$ dependence can be explained
by assuming that pions and baryons flow with opposite sign,
coupled with the measured baryon enhancement
at higher $p_t$ \cite{Spectra}. For example, taking linear functions~\cite{v1-pT-withPID} for
pion and baryon $v_1(p_t)$, we obtain a satisfactory description of our data (see
the solid curve in Fig.~\ref{fig:AuAu200GeVPt}) with pion $v_1$ slopes, $dv_1/dp_t$ =
$-0.18\pm 0.02$, $-0.34\pm 0.02$ and $-0.52\pm 0.04$,
and baryon $v_1$ slopes
$0.56\pm 0.12$,  $0.86\pm 0.10$ and $1.02\pm 0.12$ for centralities
 $0-5\%$, $5-40\%$ and $40-80\%$, respectively.
Note that the opposite $v_1$ slope for pions and protons, 
with the magnitude of proton slopes being larger,
in this case is consistent with calculations~\cite{Snellings:1999bt}
where the ``wiggle'' rapidity dependence of identified particles
has been predicted to result from the interplay of stopping and radial flow.
Currently, we are unable to test the wiggle effect in $v_1(y)$ with 
identified particles due to limited statistics and limited particle 
identification.

\begin{figure}[t]
  \includegraphics[width=.45\textwidth]{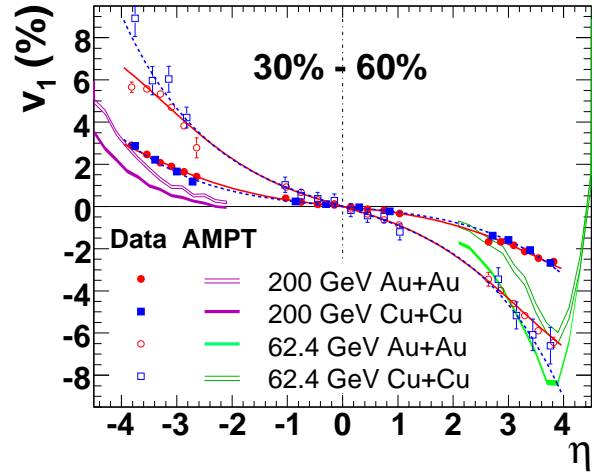}
  \caption{(color online) Charged particle $v_1(\eta)$ for mid-central 
          ($30 - 60\%$) Au+Au and Cu+Cu at 200 GeV and 62.4 GeV. 
	   The solid and dashed curves are 
           odd-order polynomial fits, to guide the eye and demonstrate the
forward-backward symmetry of the data.
	The wider shaded bands are from AMPT 
           for the same conditions as the data.  
	For clarity, 200 (62.4) GeV calculations are
	shown only at negative (positive) $\eta$.  
      The plotted error bars are statistical, and systematic errors (see 
      Figs.~\ref{fig:AuAu200GeVEta} and \ref{fig:v1_LimFr}) are within 10\%.
	  } 
           \label{fig:v1_3060Eta}
\end{figure}

\begin{figure}[t]         
  \includegraphics[width=0.450\textwidth]{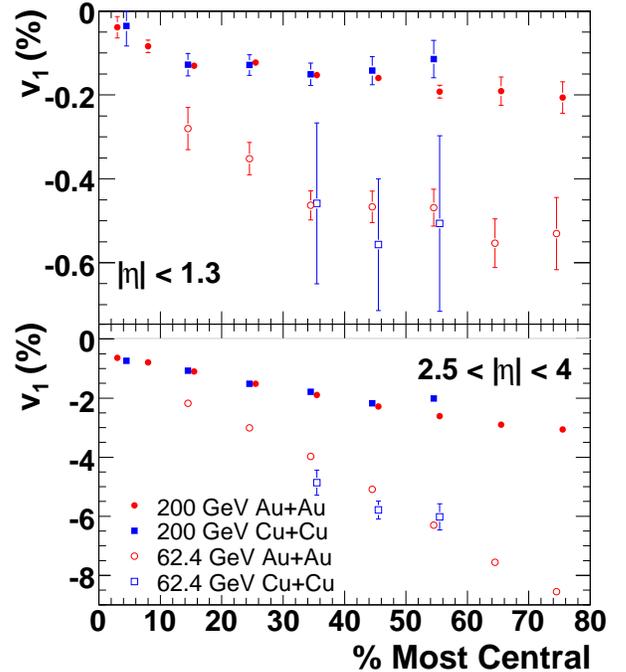}      
  \caption{(color online) Charged particle $v_1$ versus  
           centrality, for Au+Au and Cu+Cu at 200 GeV and 62.4 GeV. 
           The upper (lower) panels show results from the main TPC (FTPC).   
           The plotted error bars are statistical, and systematic errors (see 
           Figs.~\ref{fig:AuAu200GeVEta} and \ref{fig:v1_LimFr}) are within 10\%.
           } \label{fig:v1Cent}  
\end{figure}

To study the energy and system size dependence of $v_1$,
Fig.~\ref{fig:v1_3060Eta} shows Cu+Cu data compared to Au+Au 
in the centrality range  $30 - 60\%$ for both 200 and 62.4 GeV.
There is a clear trend for $v_1(\eta)$ to decrease with increasing beam energy 
for both Au+Au and Cu+Cu.  In the studied pseudorapidity and centrality range, 
$v_1(\eta)$ is, within errors, independent of the system size at each beam energy, 
despite the three-to-one mass ratio between gold and copper.  This remarkable 
feature holds for almost all centrality bins studied, as shown in Fig.~\ref{fig:v1Cent},
and persists even near mid-$\eta$ (as shown in the upper panel), where 
elliptic flow ($v_2$) of charged particles in Cu+Cu is considerably lower than 
in Au+Au~\cite{v2_scaling}.
Unlike $v_2/\epsilon$, the ratio of the elliptic flow to the system initial
eccentricity, which scales with the particle density in the transverse plane 
$(1/S) dN_{ch}/dy$~\cite{Voloshin_Poskanzer_PLB} 
(also interpreted to be the mid-rapidity area 
density~\cite{dNdyOverY} or the system 
length~\cite{Aihong}), $v_1(\eta)$ at a given centrality is found to be 
independent of the system size, and varies only with the incident energy. 
The different scalings for
$v_2/\epsilon$ and $v_1$ might arise from the way in which they are
developed: to produce $v_2$, many momentum exchanges among particles must
occur (and the number of momentum exchanges is related to the participant 
density and the dimensions of the system), while to produce $v_1$, 
an important feature of the collision process is that 
different rapidity losses need to occur (related to the incident energy) for 
particles at different distances from the center of the participant 
zone~\cite{Snellings:1999bt}.   


The hybrid transport model AMPT~\cite{ampt} lies consistently below 
the measured data, as evident from Fig.~\ref{fig:v1_3060Eta}.
STAR's prior $v_1$ study~\cite{v1ZDCSMD1} in Au+Au at 62 GeV also showed this trend 
for AMPT and other transport models. 
It is noteworthy that AMPT does not exhibit the observed pattern of system-size 
independence.  UrQMD \cite{urqmd} (not shown here) is 
similar to AMPT in exhibiting a significant change in $v_1$ between Au+Au 
and Cu+Cu. 

\begin{figure}[t]
  \includegraphics[width=0.45\textwidth]{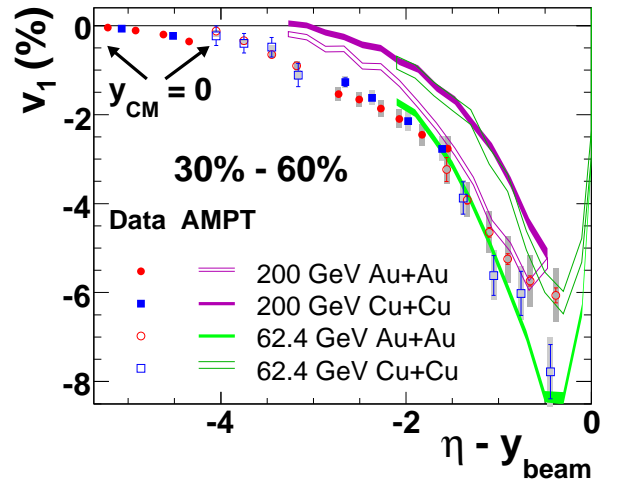}
  \caption{(color online) Charged particle $v_1$ versus $\eta - y_\mathrm{beam}$, 
           for $30 - 60\%$ Au+Au and Cu+Cu at 200 and 62.4 GeV. 
           The plotted error bars are statistical, and the shaded bars 
           show systematic errors.
           } \label{fig:v1_LimFr}
\end{figure}

Further scaling behavior is seen by transforming the data presented
in  Fig.~\ref{fig:v1_3060Eta}
into the projectile frame (see Fig.~\ref{fig:v1_LimFr}), where zero on the horizontal 
axis corresponds to the beam rapidity, $y_{\rm beam}$, for each of the collision energies.   
Within three units from $y_{\rm beam}$, most data points lie on a universal curve for 
$v_1$ versus $\eta - y_{\rm beam}$.  This incident-energy scaling of
directed flow has previously been reported for Au+Au~\cite{v1ZDCSMD1,PHOBOSv1}, and 
it is now evident that the limiting fragmentation hypothesis~\cite{LF} holds even for
much lighter collision systems like Cu+Cu.  AMPT adheres less closely to limiting
fragmentation for Cu+Cu.
Note that the quantity $\eta - y_{\rm beam}$ introduces some uncertainty due to
the use of $\eta$ instead of rapidity; the latter requires particle identification.
The system-size independence at a given fractional cross section and 
longitudinal scaling of scaled multiplicity distributions, $dN_{ch}/d\eta /(N_{part}/2)$,
have been previously reported by the PHOBOS Collaboration~\cite{:2007we}.

In summary, we have presented measurements of charged-particle 
directed flow as a function of $p_t$, $\eta$ 
and centrality in Au+Au and Cu+Cu collisions at $\sNN = $ 200 and 62.4 GeV.  
The observed trend of decreasing $v_1$ with increasing beam 
energy agrees with models.  
The lack of system-size dependence in $v_1$ for Au+Au 
and Cu+Cu is quite remarkable and is a feature not 
observed or predicted by any existing model implementation.  
The presented $\eta$-dependence of $v_1$ provides further support for 
limiting fragmentation scaling by extending its  
applicability to Cu+Cu.  
The observed $p_t$-dependence 
of directed flow motivates further theoretical investigations
and experimental measurements with identified particles.

We thank the RHIC Operations Group and RCF at BNL, and the
NERSC Center at LBNL and the resources provided by the
Open Science Grid consortium for their support. This work 
was supported in part by the Offices of NP and HEP within 
the U.S. DOE Office of Science, the U.S. NSF, the Sloan 
Foundation, the DFG Excellence Cluster EXC153 of Germany, 
CNRS/IN2P3, RA, RPL, and EMN of France, STFC and EPSRC
of the United Kingdom, FAPESP of Brazil, the Russian 
Ministry of Sci. and Tech., the NNSFC, CAS, MoST, and MoE 
of China, IRP and GA of the Czech Republic, FOM of the 
Netherlands, DAE, DST, and CSIR of the Government of India, 
Swiss NSF, the Polish State Committee for Scientific Research,
Slovak Research and Development Agency, and the Korea Sci. 
\& Eng. Foundation.
  
\end{document}